\documentclass{TeXnical/aa}
\usepackage[utf8]{inputenc}
\usepackage[varg]{txfonts}
\usepackage{physics}
\usepackage{enumitem}
\usepackage{xcolor}
\usepackage{ulem}
\usepackage{nameref}

\begin{document}
\title{Failed supernovae as a natural explanation for the binary black hole mass distribution}

\author{P. Disberg \inst{1}\thanks{paul.disberg@gmail.com}
 \and G. Nelemans\inst{1,2,3}
 }

\institute{Department of Astrophysics/IMAPP, Radboud University, P.O. Box 9010, 6500 GL Nijmegen, The Netherlands
\and
SRON, Netherlands Institute for Space Research, Niels Bohrweg 4, 2333 CA Leiden, The Netherlands
\and
Institute of Astronomy, KU Leuven, Celestijnenlaan 200D, B-3001 Leuven, Belgium}

\date{\today}

\abstract{The more gravitational wave sources are detected, the better the mass distribution of binary black holes
(BBHs) becomes known. This \textquotedblleft stellar graveyard\textquotedblright\ shows several features, including an apparent mass gap which makes the distribution bimodal. The observed chirp mass distribution, in turn, appears to be trimodal.}{We aim to investigate to which extend we can explain the observed mass distribution with stellar evolution, specifically with the hypothesis that the mass gap is caused by the difference between successful and failed supernovae (SNe).}{We pose a hypothetical remnant function, based on literature of stellar evolution simulations, which relates initial mass to remnant mass, includes a \textquotedblleft black hole island\textquotedblright\ and produces a bimodal remnant distribution. Moreover, we look at observed type II SN rates in an attempt to detect the effect of failed SNe. Finally, using a simplified estimation of binary evolution, we determine the remnant distribution resulting from our remnant function and compare it to observation.}{We find that failed SNe lower type II SN rates by approximately $25\%$, but the inferred rate from SN surveys is not accurate enough to confirm this. Furthermore, our estimation based on the remnant function produces a mass distribution that matches the general shape of the observed distributions of individual as well as chirp masses.}{Based on our research, we conclude that the failed SNe mechanism and the presence of the black hole island are a natural hypothesis for explaining the individual BBH mass distribution and chirp mass distribution. However, for a more firm conclusion more detailed simulations are needed.}
\maketitle

\section{Introduction}
\label{sec1}
\begin{figure*}
    \centering
    \includegraphics[width=17cm]{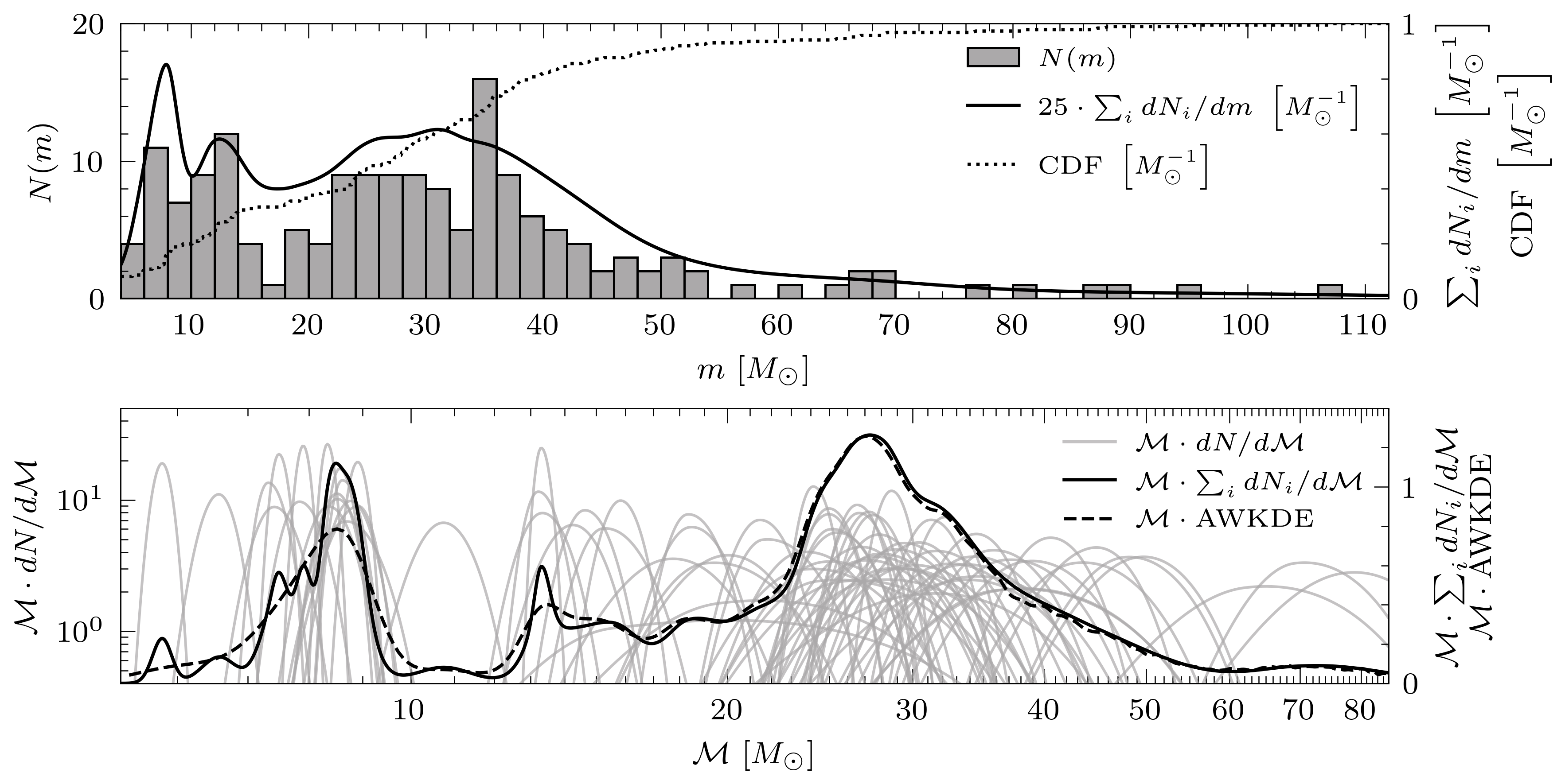}
    \caption{GW data from \citet{LIGOopensource}. The top panel shows the mass distribution of the individual BBHs (represented by a histogram of the mean values, with a bin-width of $2 M_{\odot}$). We can also represent the datapoints as asymmetric Gaussians, as described in appendix \ref{app.A}. The sum of these Gaussians, $\sum_idN_i/dm$, is given as well (solid line), together with a cumulative distribution function (CDF, dotted line on a different axis). The bottom panel shows the chirp mass distribution (grey lines) and is made to resemble the distribution from \citet{ligomassverdeling}. In order to do this, we use asymmetric Gaussians as well, which take into account the $90\%$ credible interval. We also show an adjustable-width kernel density estimation (AWKDE, dashed line), from \citet{AWKDEpaper,AWKDEcode}. The Gaussians are plotted on a logarithmic axis, while the sum and the AWKDE use a linear axis. Similarly to \citet{ligomassverdeling}, we omit  GW190814 from our analysis.}
    \label{fig1}
\end{figure*}
\noindent The more gravitational wave (GW) sources are detected, the better the distribution of the masses of the detected remnants becomes known \citep[e.g.][]{Abbott_2021}. These remnants are often black holes (BHs) in a binary system, which merge with each other after their binary evolution, through the emission of GWs \citep[e.g.][]{Abbott_2016}. This evolution includes the collapse into BHs, potentially accompanied by supernovae (SNe) of both stars, which means the SN mechanism can be important in determining the final remnant mass. We are interested in the distribution of these final masses.\\
\indent The top panel of fig. \ref{fig1} shows the distribution of the individual masses of the detected GW sources, from GWTC 1, 2 and 3 \citep{LIGOopensource}. The binary black holes (BBHs) seem to show a bimodal distribution, with a gap between $14M_{\odot}$ and $22M_{\odot}$, and an additional peak at $35M_{\odot}$, although this peak seems to disappear when taking into account the uncertainty of the data. The bottom panel shows the distribution of the corresponding chirp masses of the black hole binaries (BHBs), represented by asymmetric Gaussians (as described in appendix \ref{app.A}). This distribution appears to be trimodal \citep{ligomassverdeling}, with one mode below $10M_{\odot}$, one between $10M_{\odot}$ and $20M_{\odot}$, and one above $20M_{\odot}$. Here, there is a gap as well: at $11M_{\odot}$.\\ 
\indent One hypothethical explanation for this is given by \citet{lensing2,lensing1}, who theorize that the gap, specifically the gap in the individual mass distribution, is real and caused by gravitational lensing, because of which the distance to some of the GW sources is underestimated and therefore the chirp mass overestimated. The gap is then located between the non-lensed and lensed sources. In order to produce a distribution similar to the top panel of fig. \ref{fig1}, \citet{lensing2,lensing1} pose a merger rate which has high values at high redshift and low values at low redshift, since this produces the desired number of lensed and unlensed sources. We estimate, however, that the merger rate value they pose at high redshift implies a merger fraction which is approximately $50$ times higher than the value \citet{drakelike} estimate through their Drake-like approach. Even if we make optimistic assumptions about some of the factors in this Drake-like equation, we still find a factor of $14$ unaccounted for (as shown in appendix \ref{app.B}). Although this factor is not large enough to completely dismiss the lensing hypothesis of \citet{lensing2,lensing1}, we find it sufficient to deem their hypothesis improbable.\\
\indent We aim to explore whether the supernova (SN) mechanism could provide an alternative, more natural explanation for the observed gap. Literature suggests that the shock-wave in a core-collapse SN (CCSN) can be stalled and not cause a successful explosion \citep{mazurek1982energetics}. The star can then collapse in its totality and form a stellar remnant in a direct collapse, or \textquotedblleft failed\textquotedblright, SN. Failed SNe are difficult to detect, since they do not cause a bright explosion but instead make a massive star seemingly disappear \citep{kochanek2008survey}. Because of these difficulties, the properties of failed SNe are not well-known, even though there are several potential candidates \citep[e.g.][]{fsnsurvey5,fsnsurvey3,fsnsurvey4,fsnsurvey2,fsnsurvey1}. In addition to the search for failed SNe, others have also looked at the implications for remnant mass distributions, e.g. \citet{fsndistr2,fsndistr1} who attempts to estimate the low-mass BH distribution based on the failed SN mechanism. Overall, we state that the difference between failed SNe and successful ones could potentially create a gap in the mass distribution: some sources have a successful SN and lose mass before forming a remnant and other sources have a failed SN, conserve mass and form a more massive remnant. This means that the parameter which is important for the remnant mass distribution is the mass limit above which SNe fail.\\
\indent In this work we investigate the SN mechanism in close binaries and how it affects the BBH mass distribution (when we mention SNe we refer to CCSNe, unless specified otherwise). We start by investigating this SN mechanism in section \ref{sec2}, wherein we describe the transitions between successful and failed SNe through an initial-final mass relation based on the works of \citet{schneider} and \citet{ppisn} (section \ref{sec2.1}). We also attempt to compare this relation to observed SN rates (section \ref{sec2.2}). Then, in section \ref{sec3}, we show that this model naturally produces a bimodal mass distribution and a trimodal chirp mass distribution, similar to fig. \ref{fig1}. While our paper was under review, \citet{Schneider_2023} published a paper with a similar approach. In the discussion (section \ref{sec4}) we will make a comparison. Finally, in section \ref{sec5}, we will summarize our conclusions.  
\section{Failed supernovae}
\label{sec2}
\subsection{Remnant function}
\label{sec2.1}
\begin{figure*}
    \centering
    \includegraphics[width=17cm]{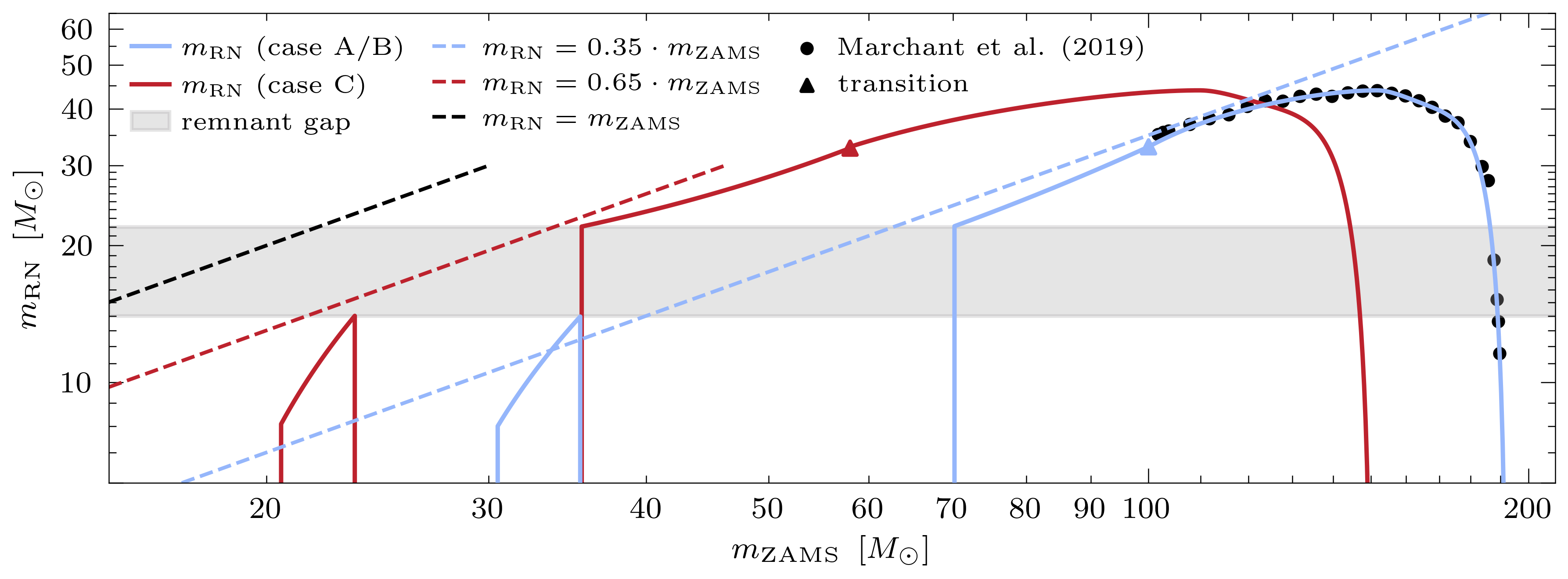}
    \caption{Remnant masses ($m_{\text{RN}}$), as function of ZAMS mass ($m_{\text{ZAMS}}$). The blue curve shows the remnant function for case A and case B mass transfer, based on the simulations by \citet{schneider} and \citet{ppisn}. The \citet{ppisn} results are shown as well (black dots). The triangle shows the transition point between the part of our function which is based on \citet{schneider} and the part which is based on \cite{ppisn}. The red curve shows the corresponding remnant function for case C mass transfer, where the part after the transition point is the PPISN fit translated to lower masses by $42M_{\odot}$. The dashed lines show $m_{\text{RN}}/m_{\text{ZAMS}}=1,\ 0.65$ and $0.35$, since the latter two are used as approximation in estimating the amount of mass transfer in our estimation. The shaded area shows the values of the remnant mass gap. Furthermore, we note that this remnant function is limited to BHs, for our purposes we are not interested in the neutron star (NS) masses. The remnant functions are explicitly given in appendix \ref{app.C}.} 
    \label{fig2}
\end{figure*}
\noindent We are interested in the differences between successful and failed SNe, and the relation between zero-age main-sequence (ZAMS) mass and the masses of the remnants they form. SN explodability simulations indicate that there may not be only one mass range for successful SNe and one range for failed SNe \citep{massrange3,massrange2,massrange1}. Instead, they find that there may be a failed SN range somewhere between $20M_{\odot}$ and $24M_{\odot}$, above which stars can have a successful SN again up until about $27M_{\odot}$, where the failed SNe take over again, supported by other studies of pre-SN compactness \citep[e.g.][]{Sukhbold_2014,Sukhbold_2016}. We combine this with the results of \citet{schneider}, who perform a simulation and investigate stellar evolution and remnant masses. They also find that there are two ranges for failed SNe: one small range for lower masses, which they call the \textquotedblleft BH island\textquotedblright, and a larger failed SN range for higher masses. They note that the two points of transition from successful to failed SNe coincide with the masses at which the stellar core change from convective to radiative carbon and neon burning, respectively. This could explain why there is a BH island in the first place. \citet{schneider} also note that their relation between ZAMS mass and remnant mass gives rise to a bimodal mass distribution.\\
\indent We construct a possible remnant function, starting by describing the BH island, based on \citet{schneider}, both for their models which concern case A/B mass transfer as well as their case C models. They describe the BH island as a function which produces remnants between approximately $8M_{\odot}$ and $10M_{\odot}$. In order to reproduce the observed BBH distribution, however, we change this to a linear function which produces remnants between $8M_{\odot}$ and $14M_{\odot}$. We give the exact definition of our remnant function, including these mass ranges, in appendix \ref{app.C}. For the other failed SNe, we use the fits given by \citet{schneider}, but because these fits start at a remnant mass of about $16M_{\odot}$, we shift them upwards to $22M_{\odot}$ instead. We motivate this by general uncertainty in the models plus the fact that rather than complete collapse or complete ejection, there could be partial fallback, so that some stars which are in the successful SN domain will in fact be able to form BHs. Finally, at the highest masses, we include the effect of pair-instability SNe (PISN) that leave behind no stellar remnant \citep{fraley1968} and pulsational pair-instability supernova (PPISN) \citep{woosley2007,woosley2017}, at slightly lower masses that can form the transition between the remnants of failed SNe and the remnantless PISNe. For this, we make a polynomial fit to the PPISN results from \citet{ppisn}. In order to create one coherent remnant function, we increase the slope of the case A/B function, connecting it to this PPISN fit, and also shift the PPISN fit for case C to lower ZAMS masses, connecting it to the case C function. The first adjustment does not influence our results significantly and the second adjustment is justified because we expect case C stars to have a more massive helium core at the end of their evolution than case A/B stars have. This is due to the fact that case C stars are able to grow more massive cores, meaning we expect case C stars to have a lower threshold for PPISNe. We show our complete remnant functions, one for case A/B and one for case C, in fig. \ref{fig2}. Here, we neglect influences of aspects such as metallicity, which we discuss in section \ref{sec4}. Nevertheless, this remnant function suffices for our purpose: showing that the bimodal BBH distribution is to be expected based on stellar evolution (which we do in section \ref{sec3}).
\subsection{Supernova rates}
\label{sec2.2}
\begin{figure}
    \resizebox{\hsize}{!}{\includegraphics{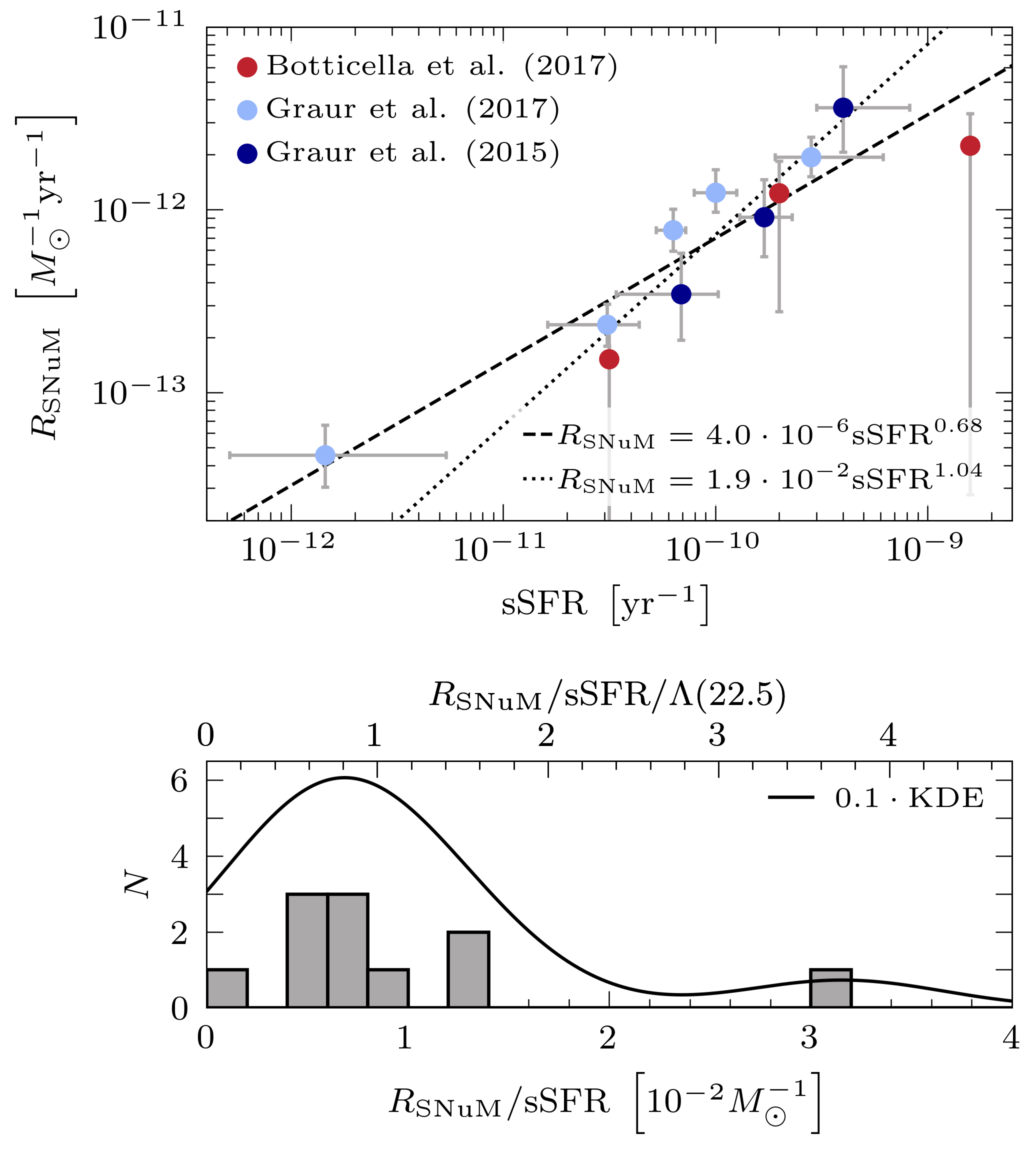}}
    \caption{Data from SN surveys \citep{Graur2015,Graur2017,Botticella2017}. The top panel shows the type II SN rates per unit mass ($R_{\text{SNuM}}$) versus the SFR per unit mass ($\text{sSFR}$), together with a fit for all the data-points (dashed line) and a fit which excludes the two outliers at $\text{sSFR}\approx1.5\cdot10^{-12}\text{yr}^{-1}\text{ and }1.5\cdot10^{-9}\text{yr}^{-1}$ (dotted line). The fits use the least-squares method on a first degree polynomial in logarithmic space and the error bars are one standard deviation. These data are corrected for the \citet{Chabrier2003} IMF, since the original data follows the \citet{salpeter} IMF. For a constant SFR, $L=\kappa\psi\int\tau\phi L(m)dm$, where $L(m)\propto m^{3.5}$ is the luminosity of a star with mass $m$. However, since $\tau(m)\cdot L(m)\propto m$, it cancels out with $\kappa$ and makes the SFR independent of the IMF. However, the SN rate is proportional to $\kappa$, so it is multiplied by a factor $\kappa_{\text{Chabrier}}/\kappa_{\text{Salpeter}}\approx1.38$, meaning that this correction also multiplies $R_{\text{SNuM}}/\text{sSFR}$ by a factor of $1.38$. The bottom panel shows a histogram of the $R_{\text{SNuM}}/\text{sSFR}$ ratio (with a bin-width of $2\cdot10^{-3}M_{\odot}^{-1}$), and a kernel density estimation (KDE).}
    \label{fig3}
\end{figure}
\noindent Before we turn to the mass distribution resulting from the remnant function, we try to detect failed SNe by investigating how they reduce SN rates. After all, since failed SNe do not produce the bright signal successful SNe do, an increase in amount of failed SNe would reduce the total SN rate. In order to make a rough estimate of the effect of failed SNe on SN rates, we assume one mass range of stars which have a successful SN, from $m_l$ to $m_u$. Above $m_u$, stars are too heavy to explode and have a failed SN, meaning they do not contribute to the SN rate. This model differs from our remnant function, since here we assume only one mass range of failed SNe, but for our purposes this suffices. In this estimation, we neglect the influences of binary interaction, effectively limiting our scope to type II SNe.\\
\indent We start by describing the SN rates. Using the initial mass function (IMF), $N(m)dm=\kappa M\phi(m)dm$ where $\kappa^{-1}=\int m\phi(m)dm$ and $\phi(m)$ the \citet{Chabrier2003} IMF shape, we express the total mass $M$ in terms of the star formation rate (SFR) $\psi(t)$. This comes down to the total mass of the stars of mass $m$ which are created a lifetime $\tau$ ago and equals $M=\psi(t-\tau(m))dt$. Therefore, the SN rate is simply the amount of stars within a certain mass range which reach the end of their stellar lifetime at time $t$ and is defined as:
\begin{equation}
    \label{eq1}
    R_{\text{SN}}(t,m_l,m_u)=\kappa\int \dfrac{M}{dt}\phi(m)dm=\kappa\int_{m_l}^{m_u}\psi(t-\tau)\phi(m)dm\quad,
\end{equation}
where we use $\tau=\tau(m)=10^{10}\text{yr}\left(m/M_{\odot}\right)^{-2.5}$. Since the precise shape of the SFR is often not determined accurately, it is difficult to determine $R_{\text{SN}}$ using this equation. Because of this, we look at the ratio between the SN rate and the SFR. The integrand of this ratio has a factor $\psi(t-\tau)/\psi(t)$, which is why we make the approximation $\psi(t-\tau)/\psi(t)\approx1$. This approximation holds for constant SFRs and decreases in accuracy for SFRs which evolve on a short timescale. Also, we set $m_l=8M_{\odot}$, which is non-controversial \citep[e.g.][]{kochanek2008survey,fsndistr1}. The ratio of SN rate to SFR becomes, then:
\begin{equation}
    \label{eq2}
    \dfrac{R_{\text{SN}}(t,8M_{\odot},m_u)}{\psi(t)}\approx\dfrac{\int_{8M_{\odot}}^{m_u} \phi(m)dm}{\int m\phi(m)dm}\equiv\Lambda(m_u)\quad,
\end{equation}
where we used the definition of $\kappa$ and introduce $\Lambda(m_u)$ as shorthand for this function. The $\kappa$ integral is taken from $m_{min}=10^{-1}M_{\odot}$ up to $m_{\max}=10^2M_{\odot}$, which means the value which does not take failed SNe into account is simply $\Lambda(m_{\max})$.\\
\indent In order to determine $m_u$, we consider the SN simulations of \citet{massrange3}, \citet{massrange2} and \citet{massrange1}, together with our remnant function based on \citet{schneider}. These simulations use different models of neutrino engines, which represent the collapsed core, and determine the mass dependent explodability per model. \citet{massrange2} find a mass dependent probability distribution, where approximately $m<20.5M_{\odot}$ and $23.5M_{\odot}<m<27M_{\odot}$ have a high probability for a successful SN. The simulations of \citet{massrange3} and \citet{massrange1} find a similar distribution, with approximately $m<21.5M_{\odot}$ and $25M_{\odot}<m<27.5M_{\odot}$ which go SN. These are in good agreement with \citet{schneider}, who find that their case C results (as shown in fig. \ref{fig2}) are similar to the results for single stars. Keeping in mind the IMF, we make a crude approximation and estimate, based on these simulations, that $m_u\approx22.5M_{\odot}$. This is comparable to the estimation of $20M_{\odot}$ which \citet{Mashian_2017} use, for instance.\\
\indent The value of $m_u$ gives $\Lambda(22.5M_{\odot})=9.06\cdot10^{-3}M_{\odot}^{-1}$. In contrast, if we neglect the influence of failed SNe, this value becomes $\Lambda(m_{\max})=1.18\cdot10^{-2}M_{\odot}^{-1}$. This means we have found a failed SNe correction of $1-\Lambda(22.5)/\Lambda(m_{\max})=0.232\approx25\%$.\\
\indent We now turn to SN surveys and use the works of \citet{Graur2015,Graur2017} and \citet{Botticella2017} in order to detect the predicted effect of failed SNe. As shown in the top panel of fig. \ref{fig3}: these data consist of the type II SN rate per unit mass ($R_{\text{SNuM}}$) versus the specific SFR ($\text{sSFR}$), i.e. the SFR per unit mass. The figure includes two fits, shaped as $R_{\text{SNuM}}=a\cdot\text{sSFR}^{b}$, where the first one includes all the data-points and the second one neglects two outliers. The fitted values of $b$, $0.68$ and $1.04$, confirm that our assumption $\psi(t-\tau)/\psi(t)\approx1$ (which means $R_{\text{SN}}\propto\psi$ and $b=1$) is appropriate. If we we assume $b\approx1$, the two points which are neglected for the second fit indeed appear to be outliers. The consequence of this assumption is that $a=R_{\text{SNuM}}/\text{sSFR}=\Lambda$. Our estimate $\Lambda(22.5M_{\odot})=9.06\cdot10^{-3}M_{\odot}^{-1}$ differs significantly from the fitted values of $a$: $4.0\cdot10^{-6}M_{\odot}^{-1}$ and $1.9\cdot10^{-2}M_{\odot}^{-1}$. If we set $b=1$, the fitted values of $a$ become $\left(7.2\pm1.7\right)\cdot10^{-3}M_{\odot}^{-1}$ and $\left(7.3\pm0.9\right)\cdot10^{-3}M_{\odot}^{-1}$ respectively, which is consistent with $\Lambda$. The bottom panel of fig. \ref{fig3} shows the distribution of the individual values of $R_{\text{SNuM}}/\text{sSFR}$, including a kernel density estimation. The distribution shows an outlier at $3\cdot10^{-2}M_{\odot}^{-1}$, which corresponds to the outlier at $1.5\cdot10^{-12}\text{yr}^{-1}$ in the top panel which was neglected in the second fit. Also, a peak is situated at approximately $0.8\cdot10^{-2}M_{\odot}^{-1}$, which is approximately the value of the fitted $a$ with $b=1$.\\
\indent The question arises whether uncertainty in other aspects of our model can account for a similar deviation from the standard $\Lambda(m_{\max})$. This deviation can, for instance, be accomplished by choosing $m_l$ to be $9.72M_{\odot}$, which seems high compared to the literature value of $8M_{\odot}$, although not unreasonably. Furthermore, uncertainties in the IMF could also account for such a deviation. Not only does \citet{Chabrier2003} give different parameter values for binary systems (which would cause a deviating value of $\Lambda$), but there is also uncertainty in the high-mass slope of the IMF. \citet{imfslope} state that this slope has a value of $\alpha=2.35_{-0.15}^{+0.35}$. A value of $2.45$ is enough to account for the $25\%$ deviation, which is well within this range. This means that even if $\Lambda(22.5M_{\odot})$ fits the data better than $\Lambda(m_{\max})$, which the data may suggest, it would be difficult to attribute this to failed SNe. Because of this, we cannot conclude that we find the effects of failed SNe in the SN survey data and therefore cannot constrain the value of $m_u$ based on this analysis. Despite not being able to confirm the failed SNe observationally, we are interested in how the remnant function from section \ref{sec2.1} shapes the BBH mass distribution.
\section{Binary black hole mass distribution}
\label{sec3}
\begin{figure*}
    \centering
    \includegraphics[width=17cm]{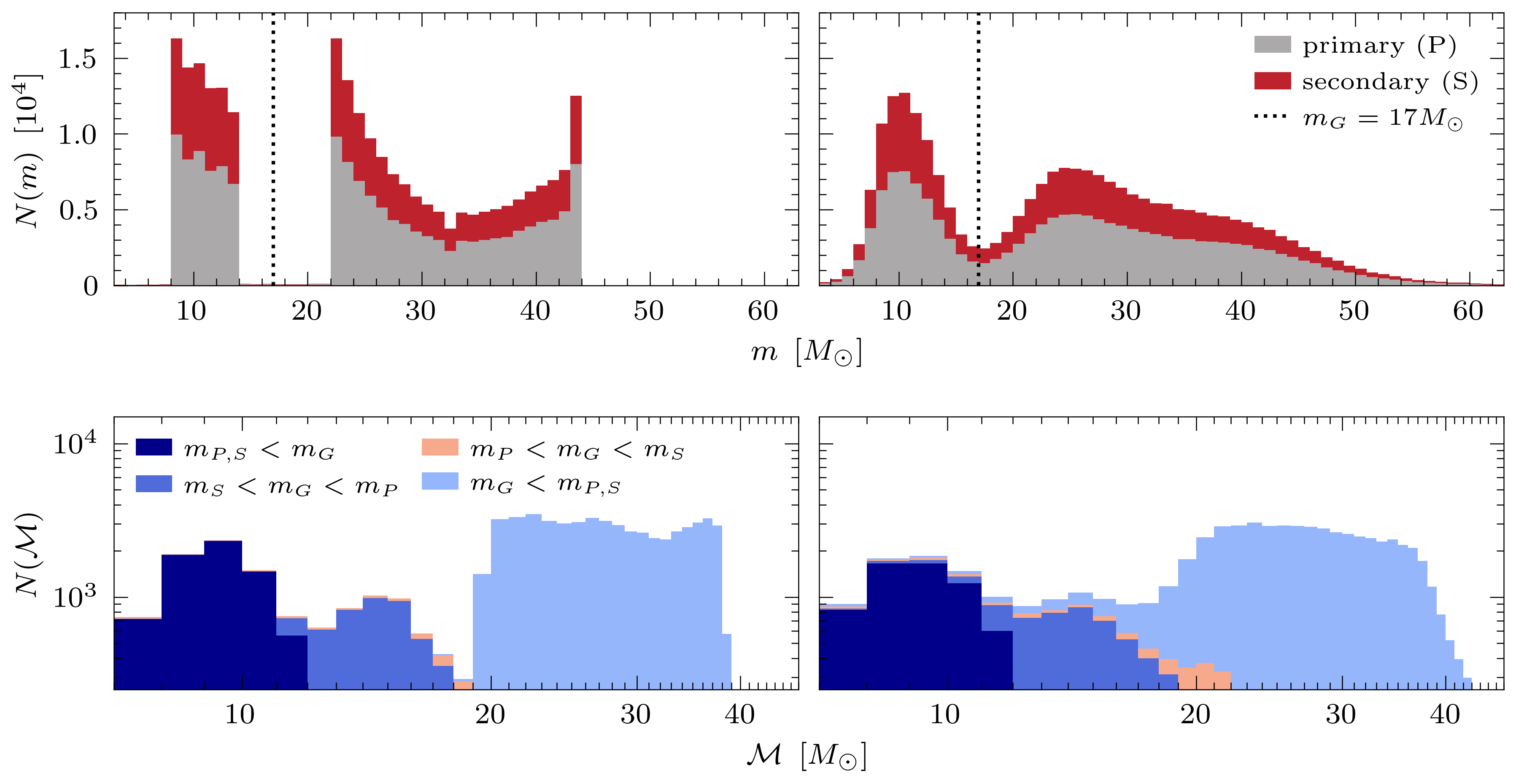}
    \caption{BBH mass distribution resulting from our estimation, where we simulate approximately $N=1.2\cdot10^6$ binaries and set $\eta=0.5$ and $\zeta=0.5$. The latter means that half of the binaries follow the case A/B remnant function, and the other half follow the case C remnant curve (as shown in fig. \ref{fig2}). The top row shows the resulting mass distribution of the individual primary (red) and secondary (grey) masses, in a stacked histogram with a bin-width of $1M_{\odot}$. The bottom row shows the chirp mass distribution in a similar histogram, where we distinguish between four possible configurations of the primary and secondary mass with respect to the central value of the mass gap ($m_G$): either both BHs are below the gap (dark blue), the secondary is below the gap while the primary is above it (blue) or vice versa (beige), or both are above the gap (light blue). The left column shows the exact results of our estimation, while the right column adds a Gaussian uncertainty to the results in order to make the simulated distribution look more similar to the observed one (as shown in fig. \ref{fig1}). We base the Gaussian uncertainty on the average standard deviations of the GW data, as described in appendix \ref{app.D}.} 
    \label{fig4}
\end{figure*}
We now want to show that our remnant function indeed produces a BBH distribution similar to observation. In order to do this, we make a simple estimation of binary evolution and determine the resulting remnant mass distribution. In our estimation, we start with approximately $1.2\cdot10^6$ binaries, in which the primary masses are distributed according to the \citet{Chabrier2003} IMF and the secondary masses are distributed uniformly. This means that we define the primary star here to have the highest \textit{initial} mass.\\
\indent In our model, we also include mass transfer from the primary to the secondary in a Roche-lobe overflow (RLO) phase, and neglect mass transfer from the secondary to the primary. We state that the secondary accretes a fraction $\eta$ of the mass expelled from the primary during the RLO phase. The precise value of $\eta$ is uncertain \citep{dorozsmai}, so we will simply set $\eta=0.5$, since it does not affect the general shape of the remnant distribution. In order to determine the amount of mass which is expelled from the primary, we cannot simply take the difference between ZAMS mass and remnant mass, since in successful SNe there is mass loss outside of the RLO phase. We therefore approximate the expelled mass of the primary in the RLO phase by $0.35\cdot m_{\text{ZAMS}}$ for case A/B and $0.65\cdot m_{\text{ZAMS}}$ for case C, since these curves approximate the failed SN curves, as shown in fig. \ref{fig2}. The secondary, then, accretes a fraction $\eta$ of this expelled mass and is rejuvenated, which means that we use this new mass as if it were the ZAMS mass. Besides $\eta$ we also define the parameter $\zeta$, which equals the fraction of binaries which follow the case C curve. Since $\zeta$ also does not influence the general shape of the distribution significantly, we set $\zeta=0.5$ as well.\\
\indent The estimated mass distribution is shown in fig. \ref{fig4}. The left panels show the exact results from our estimation and the right panels show the results with an added Gaussian uncertainty, as described in appendix \ref{app.D}. Unsurprisingly, the remnant mass distribution is indeed bimodal: it shows one peak at lower masses, caused by the BH island, and one peak at higher masses. The exact results of our estimation, without the Gaussian uncertainty, also show a PPISN peak around $44M_{\odot}$, caused by the flat PPISN distribution in fig. \ref{fig2}. The general shape of the estimated distribution is similar to the observed distribution, although it is difficult to compare the two in detail because the relative heights of the peaks are influenced by multiple aspects. Not only is there a bias towards high-mass mergers in observation, but the parameters $\eta$ and $\zeta$ also influence the heights. The effects of these parameters, including the fact that varying them does not influence our conclusions about the general distribution, is shown in appendix \ref{app.E}.\\
\indent The resulting chirp mass distribution also shows interesting features. Since the individual mass distribution is bimodal, with one peak at either side of the mass gap at $m_G=17M_{\odot}$, the chirp mass distribution shows four possible configurations, with the primary and secondary mass at the same and opposite sides of the mass gap. These configurations can clearly be found in the chirp mass distribution: there is one peak centered just below $10M_{\odot}$ for binaries with both BHs below the mass gap, one peak centered around $28M_{\odot}$ for binaries with both BHs above the mass gap, and also two peaks centered around $15M_{\odot}$, representing a mixed population with BHs at either side of the gap. The latter mostly consists of binaries where the primary is above the gap and the secondary below, although, depending on $\eta$, a small population of binaries where only the secondary is above the gap can be found as well. These features can be linked to the observed chirp mass distribution, because the observed distribution (fig. \ref{fig1}) shows a trimodal distribution very similar to our results: one peak just below $10M_{\odot}$, one around $28M_{\odot}$, a small group of binaries around $15M_{\odot}$ and a gap just above $10M_{\odot}$. We also predict a gap just below $20M_{\odot}$, which is somewhat visible in our AWKDE but even more visible in the estimation by \citet{ligomassverdeling}. Our results are not identical to the observed chirp mass distribution, especially above $20M_{\odot}$ they seem to differ, but the general shapes of the distributions do seem to agree.\\
\begin{figure*}
    \sidecaption
    \includegraphics[width=12cm]{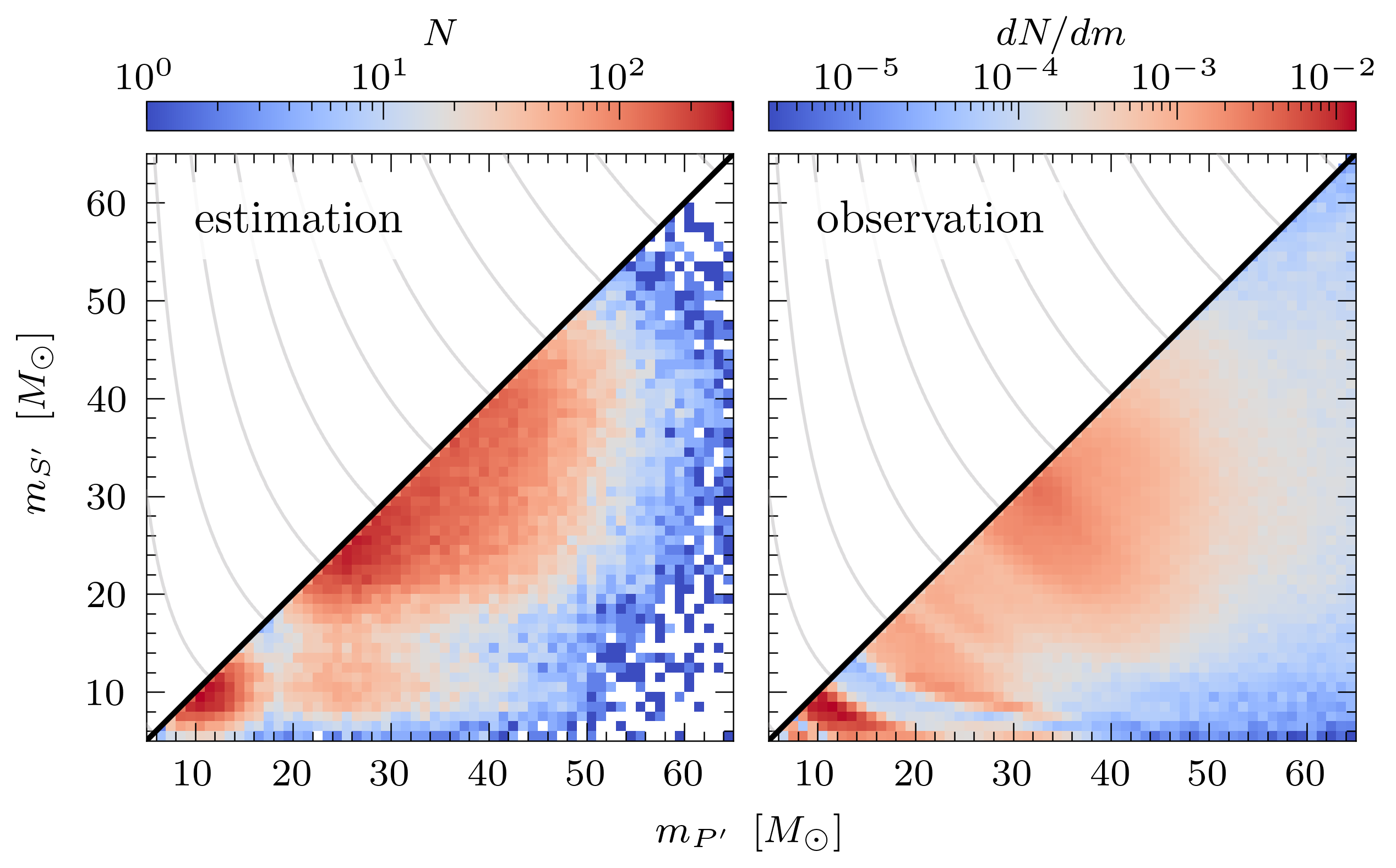}
    \caption{Primary versus secondary mass distributions, for the results of our estimation (left) and the GW data (right), both in two-dimensional histograms with a bin-width of $1M_{\odot}$. The left panel shows the upper right panel from fig. \ref{fig4}. We use the notation $P^\prime$ and $S^\prime$ here to denote the stars with the largest and smallest final masses, respectively, in contrast to $P$ and $S$ from fig. \ref{fig4}, which concern the \textit{initial} masses. The right panel shows the cosmologically reweighted posterior distributions of the GW data \citep{parameter-GWTC2,parameter-GWTC3}. Since the posteriors are not normalized, we rescaled them to $\int_{0M_{\odot}}^{100M_{\odot}}\frac{dN}{dm}dm\approx1.4\cdot10^5$. Then, after adding all the posteriors, we normalized the resulting distribution in between $5M_{\odot}$ and $65M_{\odot}$. We omit GW200322 here, since no valid posterior for this source has been given. The grey lines are lines of equal chirp mass, with intervals of $5M_{\odot}$.}    
    \label{fig5}
\end{figure*}
\indent In a comparison between our simple estimation and the GW data, we examine the primary versus secondary mass distribution. Fig. \ref{fig5} shows this distribution, together with the posterior distributions of the GW observations. Both the simulated and the observed distributions clearly show the BH island and the other failed SNe. Also, there seem to be some observed mergers which can be linked to the mixed population. There are, however, also differences between the two distributions. For example, there is a local maximum at $m_{P^\prime}\approx m_{S^{\prime}}\approx34M_{\odot}$, while the corresponding maximum in our distribution is situated around $m_{P^\prime}\approx m_{S^{\prime}}\approx28M_{\odot}$. Also, the observed distribution has relatively more BHs in the BH island, even though there is an observational bias towards high-mass mergers. However, we are interested in the general shape of the distribution and not normalisation and details. After all, consideration of partial fallback as well as the parameters $\eta$ and $\zeta$ influence the relative amount of BHs in the BH island. Another interesting feature of fig. \ref{fig5} is the fact that the observed mass gap seems to be situated at lower masses than the simulated mass gap: while we set the simulated mass gap at $m_{P^\prime}\approx m_{S^{\prime}}\approx17M_{\odot}$, the observed gap appears to be around $m_{P^\prime}\approx m_{S^{\prime}}\approx14M_{\odot}$. This could mean that we did not have to shift the results of \citet{schneider} upwards in our remnant function at all. The difference can be explained by the fact that applying the Gaussian uncertainty to the individual masses results in a two-dimensional distribution which is more symmetric than the observed posteriors. Finally, it is interesting to compare fig. \ref{fig5} with the rate densities determined by \citet{ligomassverdeling}. Where we find three populations: a BH island, other failed SNe and a mixed population, they find a similar distribution but for NS-NS, BH-BH and BH-NS mergers, respectively.
\section{Discussion}
\label{sec4}
\noindent We have proposed failed SNe, and in particular the difference between failed and successful SNe, as a natural explanation for the shape of the BBH mass distribution. In order to do this, we posed a remnant function (section \ref{sec2.1}), investigated type II SN rate data (section \ref{sec2.2}), and examined the BBH distribution caused by this function (section \ref{sec3}). Our model is rather simplistic, however, and we do not claim that our results describe the actual binary mass distribution in detail. We are therefore careful not to draw too strong conclusions in comparing our results from fig. \ref{fig4} and fig. \ref{fig5} to the data as shown in fig. \ref{fig1}. Although, the simplicity of our model does indicate that our conclusions are a robust feature of BBH distributions caused by remnant functions similar to ours.\\
\indent Furthermore, we argue that our remnant function is a more natural hypothetical explanation for the BBH mass distribution than the gravitational lensing hypothesis mentioned in the introduction, for two reasons. Firstly, although they are difficult to compare, we estimate the difference of a factor 50 between the \citet{lensing2,lensing1} merger rate and an optimistic BBH fraction estimate, discussed in appendix \ref{app.B}, to be greater than the difference between our remnant function and the results of \citet{schneider} and \citet{ppisn}. Secondly, the chirp mass distribution is difficult to explain using the lensing hypothesis. After all, the two stars in the binary are either both lensed or both non-lensed. This means that there should not have been any binary where the primary is above the gap and the secondary below. The lensing hypothesis, therefore, has trouble explaining the trimodal chirp mass distribution, while our remnant function reproduces it.\\
\indent As mentioned in the introduction: \citet{Schneider_2023} published a paper which uses a similar method and reaches comparable conclusions, while our paper was in review. They use simulations equivalent to those of \citet{schneider}, showing a BH island, which result in a bimodal distribution of the individual BBHs. Their BBH distribution looks similar to the one we find, except for the PPISN peak since they do not incorporate these. Moreover, they argue that the bimodal mass distribution causes a trimodal chirp mass distribution with the same argument we use: the mixing between BHs from the BH island and the higher mass BHs. They also note that fallback after the SN can extend the BH island to lower masses, because of the same reason we have extended our BH island in fig. \ref{fig2} to include lower masses. Furthermore, according to \citet{Schneider_2023}, metallicity mainly affects the case A/B curve in fig. \ref{fig2}, but because of the mass-loss through wind associated with metallicity they state that an increase in metallicity can shift the resulting mass distribution to lower masses. Overall, we find that their results are in good agreement with our findings, which strengthens our argument.
\section{Conclusions}
\label{sec5}
\noindent After our analysis of the BBH mass distribution and chirp mass distribution (section \ref{sec3}) resulting from our remnant function (section \ref{sec2}), we conclude the following:
\begin{itemize}
    \item The mass distribution of BBHs appears to show a gap for approximately $14M_{\odot}<m<22M_{\odot}$ (fig. \ref{fig1}). Gravitational lensing as an explanation for this gap \citep{lensing2,lensing1} seems, initially, to assume an unreasonably high merger rate for $z>2$ (fig. \ref{fig2}). We find that this merger rate implies a BBH fraction which is approximately $50$ times larger than the one described by \citet{drakelike}. This is improbably large, but not large enough to completely dismiss this possibility.
    \item We investigate failed SNe as a more natural explanation for the gap. Firstly, we approximate the results of the simulations by \citet{schneider} and \citet{ppisn} with a comprehensive remnant function which describes the relation between ZAMS mass and remnant mass, for case A/B and case C binaries (fig. \ref{fig2}). This function includes a BH island, which existence is supported by SN explodability simulations and is the cause of the bimodality in the BBH mass distribution.
    \item We investigate whether we can confirm the mass range for failed SNe observationally. We assume a single mass range for successful SNe, which turn into failed SNe above a certain mass limit. We take an optimistic value of $22.5M_{\odot}$ for this mass limit and find that this implies a $R_{\text{SN}}/\text{SFR}$ reduction of approximately $25\%$, since failed SNe are not detected and included in the type II SN rate. However, even though this reduction is compatible with SN survey data to some degree, the data is quite uncertain and small changes in other model parameters could also account for a similar reduction in SN rate. We therefore cannot state that the failed SNe mass range can be confirmed by the SN survey data.
    \item Using this remnant function, a bimodal BBH distribution can be estimated which look similar to observation. Also, our simplistic estimation produces a trimodal chirp mass distribution, which corresponds to observation. The primary versus secondary mass distribution resulting from our estimation corresponds to some degree with the posterior samples of the observed GW sources. This distribution clearly shows a mass gap, and a comparison of estimation and observation indicates that our remnant function does not need to deviate from literature values in the degree that it does, strengthening our conclusions.
\end{itemize}
\noindent Based on our results, we conclude that failed SNe, and in particular the relation between ZAMS mass and remnant mass which includes a BH island, can provide a natural explanation for the apparent bimodality in the observed BBH mass distribution. We therefore state that, when trying to explain the general shape of the observed BBH distribution, a consideration of stellar evolution is shown to be fruitful.\\
\indent Future research could expand or improve multiple aspects of our work. Reproducing the observed BBH distribution in detail, for instance, would be an interesting topic of research. Not only could the remnant function perhaps stay closer to literature values and still produce the desired mass gap, as implied by fig. \ref{fig5}, but other aspects such as a consideration of partial fallback, PPISNe and the values of $\eta$ and $\zeta$ also influence the relative heights of the different peaks. It would be an interesting endeavour to create a model which includes these aspects in such a way that it reproduces the observed distribution in more detail. An interesting question could be if the chirp mass distribution which corresponds to such a detailed model also resembles observation in detail. And, finally, more (precise) SN survey and GW data would improve the observational verification, which could help in identifying the precise effects of failed SNe.

\newpage
\begin{acknowledgements}
    This research has made use of data or software obtained from the Gravitational Wave Open Science Center (gw-openscience.org), a service of LIGO Laboratory, the LIGO Scientific Collaboration, the Virgo Collaboration, and KAGRA. LIGO Laboratory and Advanced LIGO are funded by the United States National Science Foundation (NSF) as well as the Science and Technology Facilities Council (STFC) of the United Kingdom, the Max-Planck-Society (MPS), and the State of Niedersachsen/Germany for support of the construction of Advanced LIGO and construction and operation of the GEO600 detector. Additional support for Advanced LIGO was provided by the Australian Research Council. Virgo is funded, through the European Gravitational Observatory (EGO), by the French Centre National de Recherche Scientifique (CNRS), the Italian Istituto Nazionale di Fisica Nucleare (INFN) and the Dutch Nikhef, with contributions by institutions from Belgium, Germany, Greece, Hungary, Ireland, Japan, Monaco, Poland, Portugal, Spain. The construction and operation of KAGRA are funded by Ministry of Education, Culture, Sports, Science and Technology (MEXT), and Japan Society for the Promotion of Science (JSPS), National Research Foundation (NRF) and Ministry of Science and ICT (MSIT) in Korea, Academia Sinica (AS) and the Ministry of Science and Technology (MoST) in Taiwan. This research is supported by the Netherlands Organisation for Scientific Research (NWO). We thank both Onno Pols for useful discussions and the anonymous referees who provided comments which helped to improve this paper.
\end{acknowledgements}
\bibliographystyle{TeXnical/aa}
\bibliography{References}

\begin{appendix}
\section{Asymmetric Gaussians}
\label{app.A}
\begin{figure}[h]
    \resizebox{\hsize}{!}{\includegraphics{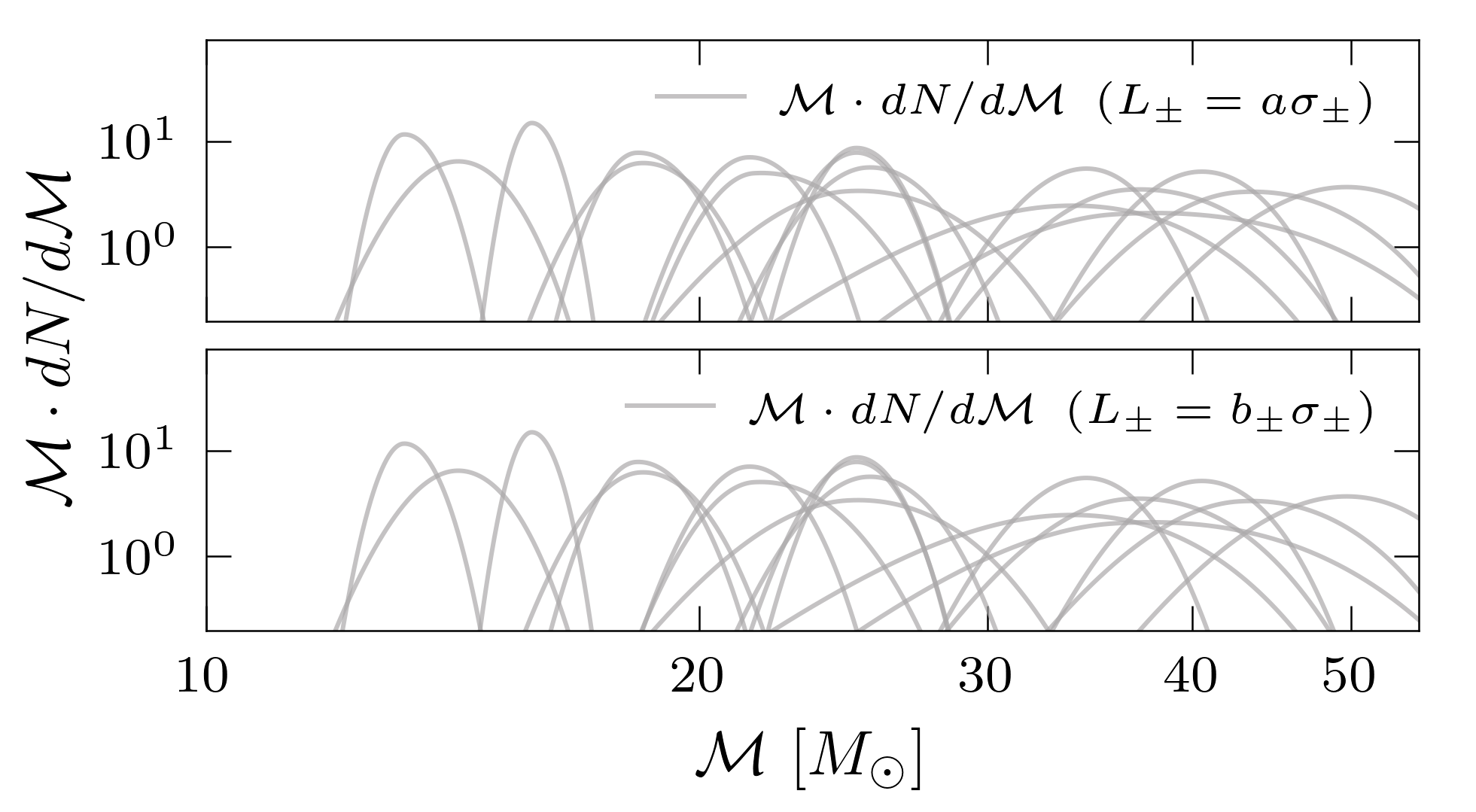}}
    \caption{A randomly selected subset of the chirp masses \citep{LIGOopensource}, represented by asymmetric Gaussians (as in fig. \ref{fig1}) which are similar to the posterior distributions of \citet{ligomassverdeling}. The $90\%$-interval is determined by either the $a$-method from eq. \ref{eqa5} (top panel) or the $b_{\pm}$-method from eq. \ref{eqa7} (bottom panel).}
    \label{figa1}
\end{figure}

\noindent In order to approximate the posterior chirp mass distribution from \citet{ligomassverdeling} in fig. \ref{fig1}, we use Gaussian probability density functions with two different standard deviations. This is equivalent to using two halves of different Gaussians which meet at the mean. These functions are described by the standard Gaussian formula:
\begin{equation}
    \label{eqa1}
    f^\pm(x)=\dfrac{1}{\sigma_\pm\sqrt{2\pi}}\exp\left(-\dfrac{1}{2}\left(\dfrac{x-\mu}{\sigma_\pm}\right)^2\right)\quad,
\end{equation}
where $f^{-}(x)$ describes the function below the mean $\mu$ with standard deviation $\sigma_-$ and $f^{+}(x)$ describes the function above the mean with standard deviation $\sigma_+$. In other words, $\sigma_{\pm}=\sigma_-$ if $x\leq\mu$ and $\sigma_{\pm}=\sigma_+$ if $x>\mu$. Also, $\overline{\sigma}$ is simply the average standard deviation $\left(\sigma_-+\sigma_+\right)/2$. The combination of these halves can then be constructed as
\begin{equation}
    \label{eqa2}
    f(x)=k_1\cdot\left\{\begin{matrix}f^-(x)\hfill&\text{for }x\leq\mu\\k_2f^+(x)&\text{for }x>\mu\end{matrix}\right.\quad,
\end{equation}
where the constant $k_2$ makes sure $f^-(\mu)=k_2f^+(\mu)$ and $k_1$ ensures normalization. From this description of $k_2$ and eq. \ref{eqa1} it immediately follows that $k_2=f^{-}(\mu)/f^{+}(\mu)=\sigma_+/\sigma_-$.
Also, the normalization requirement can be used to determine $k_1$: $\int_{-\infty}^{\infty}f(x)dx=1$ gives $k_1=\sigma_-/\overline{\sigma}$. Filling in the values of $k_1$ and $k_2$ gives, then,  the overall asymmetric Gaussian
\begin{equation}
    \label{eqa3}
    f(x)=\dfrac{1}{\overline{\sigma}\sqrt{2\pi}}\exp\left(-\dfrac{1}{2}\left(\dfrac{x-\mu}{\sigma_\pm}\right)^2\right)\quad,
\end{equation}
which has the following antiderivative:
\begin{equation}
    \label{eqa4}
    F(x)=\dfrac{\sigma_{\pm}}{2\overline{\sigma}}\erf\left(\dfrac{x-\mu}{\sqrt{2}\sigma_{\pm}}\right)\quad.
\end{equation}
\indent However, the asymmetry brings with it some ambiguity. When only $\mu$ and the $90\%$ credible deviations ($L_-$ and $L_+$) are given, it is not immediately obvious how to define a 90\% interval in equation \ref{eqa3}. As a first option, it is possible to assume that the interval, $\mu-L_-\leq x\leq\mu+L_+$, has borders proportional to the standard deviations with a constant $a$ (i.e. $L_\pm=a\sigma_\pm$). This means $\int_{\mu-a\sigma_-}^{\mu+a\sigma_+}f(x)dx=90\%$, or
\begin{equation}
    \label{eqa5}
    90\%=\erf\left(\dfrac{a}{\sqrt{2}}\right)\implies a=\sqrt{2}\erf^{-1}\left(90\%\right)\approx1.65\quad.
\end{equation}
Alternatively, it is possible to define the 90\% interval as \textquotedbl cutting of\textquotedbl\ 5\% on either side. The relation between the interval borders and the standard deviations then no longer has a single proportionality constant but two: $b_-$ and $b_+$ (i.e. $L_{\pm}=b_{\pm}\sigma_{\pm}$). The fact that $\int_{-\infty}^{\mu-L_-} f(x)dx=\int_{\mu+L_+}^{\infty}f(x)dx=5\%$ translates into
\begin{equation}
    \label{eqa6}
    5\%=\dfrac{\sigma_\pm}{2\overline{\sigma}}-\dfrac{\sigma_\pm}{2\overline{\sigma}}\erf\left(\dfrac{L_\pm}{\sqrt{2}\sigma_\pm}\right)=\dfrac{1-\erf\left(b_\pm/\sqrt{2}\right)}{1+\sigma_\mp/\sigma_\pm}\quad.
\end{equation}
This equation can be transformed, after substitution of $\sigma_{\pm}=L_{\pm}/b_{\pm}$, into the two ($\pm$) equations: 
\begin{equation}
    \label{eqa7}
    b_\pm=\sqrt{2}\erf^{-1}\left(95\%-5\%\dfrac{b_\pm}{b_\mp}\dfrac{L_\mp}{L_\pm}\right)\quad,
\end{equation}
which can be solved numerically to obtain $b_-$ and $b_+$. It is clear that eq. \ref{eqa7} satisfies $\sigma_-=\sigma_+\rightarrow b_-=b_+=a$. Fig. \ref{figa1} shows, similarly to fig. \ref{fig1}, the chirp mass distribution represented by both methods of asymmetric Gaussians (to avoid a chaotic plot, only a subset of the data is shown). It is clear that both methods, using either $a$ or $b_\pm$, give almost indistinguishable results for the data-set. In fact, the total distributions differ less than 5\%. Because of this, we use the $a$-method (eq. \ref{eqa5}) in this work, which is mathematically more straightforward.
\section{Merger rate}
\label{app.B}
\begin{figure}[h]
    \resizebox{\hsize}{!}{\includegraphics{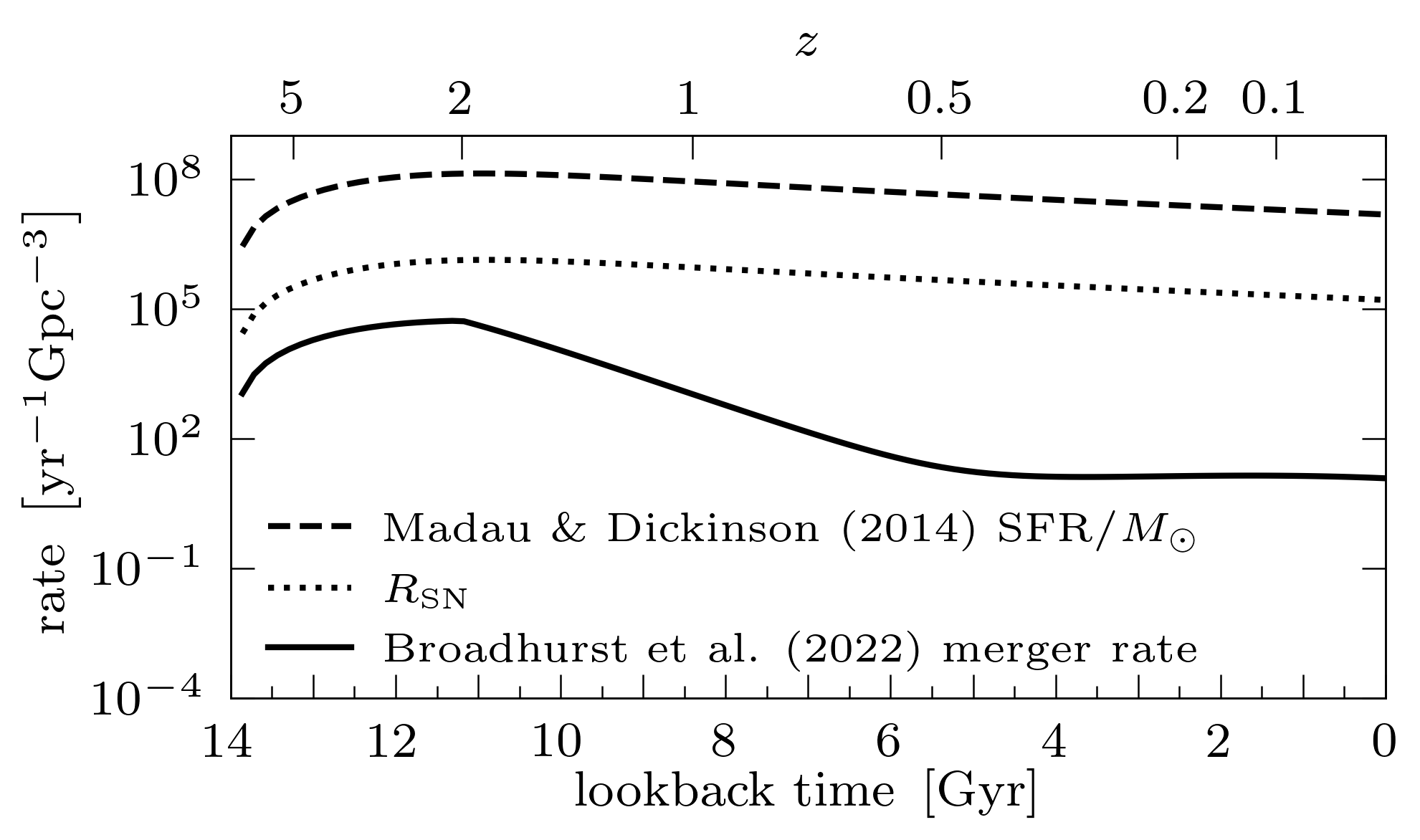}}
    \caption{\citet{MadauDickinson} star formation rate per $M_{\odot}$, the corresponding SN rate with $m_l=8M_{\odot}$ and $m_u=30M_{\odot}$ (through eq. \ref{eq1}) and the \citet{lensing2,lensing1} BBH merger rate.}
    \label{figb1}
\end{figure}

\noindent According to \citet{lensing2,lensing1}, gravitational lensing can explain the apparent bimodality in the BBH mass distribution, as shown in the top panel of fig. \ref{fig1}. They note that gravitational lensing focuses the GW signal, causing the distance of the merger to be underestimated by a factor equal to the square root of the magnification. This means that the inferred chirp mass is overestimated. After all, an increased distance means a higher redshift, or a lower frequency, which gives a higher inferred chirp mass. They state that the gap in BBH masses can be explained by the hypothesis that a significant part of the observed BBHs is gravitationally lensed. In order to explain the observed BBH mass distribution, their model needs a merger rate which has a high value at high redshift, since there has to be a significant amount of mergers to be lensed by massive objects at lower redshift. This merger rate, then, has to decline rapidly towards lower redshift. They pose this hypothetical merger rate, not as a result from a physical model, but because from it the desired bimodal BBH distribution follows. They explicitly assume all mergers with a chirp mass above $20M_{\odot}$ to be gravitationally lensed \citep{lensing1}.\\
\indent The merger rate they use takes the shape of $R_{\text{merger}}(z)=a\cdot S(z)\cdot F(z)$, where the merger rate ($R_{\text{merger}}$) is given in terms of a constant $a=6\cdot10^3\ \text{yr}^{-1}\text{Gpc}^{-3}$, as well as a function $S(z)$ which traces the \citet{MadauDickinson} SFR and a modulation function $F(z)$, which equals $1$ for $z>2$ and causes an exponential decline for $2>z>0.5$. Fig. \ref{figb1} shows both the merger rate and the \citet{MadauDickinson} SFR, together with the SN rate, which we define in section \ref{sec2.2}.\\
\indent The merger rate seems to be extremely high for $z>2$. In order to investigate if this is at all possible, we turn to the work of \citet{drakelike}, who use a \textquotedblleft Drake-like\textquotedblright\ equation to estimate the fraction of binaries that produce a merging BBH ($f_{\text{BBH}}$). In order to produce a BBH merger, the following fractions need to be multiplied: the fraction of primary stars which fall within the SN range ($f_{\text{prim}}$), the fraction of their secondary stars which fall inside this range ($f_{\text{sec}}$), the fraction of these binaries which have a separation suitable for merging ($f_{\text{init sep}}$), the fraction of these binaries which survive the primary SN ($f_{\text{survive SN1}}$) as well as the common envelope phase ($f_{\text{CE}}$) and the secondary SN ($f_{\text{survive SN2}}$), and finally the fraction of these binaries which successfully merge ($f_{\text{merge}}$). In total this gives: 
\begin{align}
\begin{split}
    \label{eqb1}
    f_{\text{BBH}}&=f_{\text{prim}}\cdot f_{\text{sec}}\cdot f_{\text{init sep}}\cdot f_{\text{survive SN1}}\cdot f_{\text{CE}}\cdot f_{\text{survive SN2}}\cdot f_{\text{merge}}\\
    &\approx 0.001\cdot0.5\cdot0.5\cdot1\cdot0.1\cdot1\cdot0.2=5\cdot10^{-6}\quad,
\end{split}
\end{align}
where we filled in the estimations of the fractions made by \citet{drakelike}.\\
\indent In order to compare the merger rate inferred by \citet{lensing2,lensing1} to eq. \ref{eq1}, this merger rate has to be converted to a BBH fraction (which we will call $f'_{\text{BBH}}$), this equals the ratio between the number of binary stars that merge and the total number of stars. The number of binary stars that merge is calculated by taking twice the merger rate (since a merger consists of two stellar remnants). The latter is determined through the SFR. However, we need to consider the fact that the SFR is determined through the observed luminosity. We define the binary to have a certain mass ratio $q=m_S/m_P$, which means that the total luminosity of the binary can be described by $L\propto m_P^{3.5}+m_S^{3.5}=\left(1+q^{3.5}\right)m_P^{3.5}$. In other words, the SFR implies $1+q^{3.5}$ stars while there are actually $2$. Therefore the total number of stars has to be adjusted by a factor $2/\left(1+q^{3.5}\right)$. Here, we make the approximation $2/\left(1+q^{3.5}\right)\approx2$, which is valid even if $q$ takes values as high as $0.6$ or $0.7$.\\
\indent We use the \citet{Chabrier2003} IMF and compute the IMF integrals from $m_{\min}=10^{-1}M_{\odot}$ to $m_{\max}=10^2M_{\odot}$. The total amount of stars is, then, simply $\kappa M\int\phi(m)dm$. If we assume a \citet{MadauDickinson} SFR, which is the function $S(z)$ multiplied by a constant $b=1.5\cdot10^7M_{\odot}\text{yr}^{-1}\text{Gpc}^{-3}$, we get for $z>2$:
\begin{align}
\begin{split}
    \label{eqb2}
    f'_{\text{BBH}}&=\dfrac{N_{\text{BBH}}}{N_{\text{tot}}}=\dfrac{2R_{\text{merger}}/dt}{2\kappa\psi(z)/dt\int\phi(m)dm}=\dfrac{aS(z)F(z)}{bS(z)}\dfrac{\int m\phi(m)dm}{\int\phi(m)dm}\\
    &\approx \dfrac{6\cdot10^3}{1.5\cdot10^7}\cdot0.67=2.7\cdot10^{-4}\quad,
\end{split}
\end{align}
where $\psi(z)$ is the SFR at redshift $z$, which means $M=\psi(z)/dt$. We do not take into account the fact that there is a time difference between the stellar formation and the merger events, effectively assuming that the bulk of the mergers occur on a relatively short timescale, compared to the SFR.\\
\indent In order to analyze $f'_{\text{BBH}}$, we use fractions similar to those used in eq. \ref{eqb1}. Leaving out $f_{\text{survive SN1}}$ and $f_{\text{survive SN2}}$, since they equal $1$, the remaining factors have to equal $f'_{\text{BBH}}=2.7\cdot10^{-4}$ and account for the deviation from $f_{\text{BBH}}$ of $f'_{\text{BBH}}/f_{\text{BBH}}=56\approx50$. The question, now, is if this is possible. Using IMF integrals, combined with the $2/\left(1+q^{3.5}\right)$ correction, we estimate that reasonable adjustments to $f_{\text{prim}}$ and $f_{\text{sec}}$ could both account for a factor of approximately $2$. This means that the remaining $f'_{\text{init sep}}$, $f'_{\text{CE}}$ and $f'_{\text{merge}}$ have to account for a factor of $56/4\approx14$, or an average factor of $14^{1/3}\approx2.4$ per fraction. This factor is big, but not large enough to completely dismiss the \citet{lensing2,lensing1} merger rate at $z>2$. After all, the factor of $14$ can be distributed between the remaining fractions without one of them becoming larger than $1$.\\
\indent In contrast, if we look at the merger rate at $z<0.5$, the fraction seems to become quite small. This is due to the factor $F(z)$ in the merger rate. If we approximate the merger rate below $z=0.5$ as a constant value (specifically as the value at $z=0.2$), we find that this value equals $F(0.2)\cdot f'_{\text{BBH}}(z>2)\approx1.4\cdot10^{-3}\cdot2.7\cdot10^{-4}=3.8\cdot10^{-7}$. This is approximately $8\%$ of the \citet{drakelike} fraction (eq. \ref{eq1}). Since $f_{\text{BBH}}$ consists of $7$ fractions, each has to be adjusted by a factor of $0.08^{1/7}\approx0.7$, on average, which is well within their uncertainty.
\section{Remnant function}
\label{app.C}
The remnant function we pose in fig. \ref{fig2} consists of the following equations. Firstly, we follow \citet{woosley2019coremass} for the relation between ZAMS mass and the final mass of the helium core:
\begin{equation}
    \label{eqc1}
    f_{\text{He}}(m)=\left\{\begin{matrix}0.0385m^{1.603}\hfill&\text{for }m\leq30M_{\odot}\hfil\\0.50m-5.87M_{\odot}\hfill&\text{for }m>30M_{\odot}\end{matrix}\right.\quad.
\end{equation}
We use this relation to rewrite the results of \citet{ppisn} as a function of ZAMS mass instead of helium core mass. This takes the form of two fourth degree polynomials, shaped as
\begin{equation}
    \label{eqc2}
    f_{\text{PPISN}}(m)=\left\{\begin{matrix}M_{\odot}\sum_{i=0}^4a_i\left(\dfrac{m}{M_{\odot}}\right)^i&\text{for }m\leq 151.74M_{\odot}\hfill\\M_{\odot}\sum_{i=0}^4b_i\left(\dfrac{m}{M_{\odot}}\right)^i&\text{for }m>151.74M_{\odot}\hfill\end{matrix}\right.\quad,
\end{equation}
where the coefficients $a_i$ are determined through a polynomial fit through the data points at $f_{\text{He}}(m)/M_{\odot}=100,115.74,127.74$ and $151.74$, and $b_i$ is given as a fit through the data points at $f_{\text{He}}(m)/M_{\odot}=151.74,167.74,179.74$ and $189.74$, as given by \citet{ppisn}. The two polynomials connect at $m=f^{-1}_{\text{He}}(70M_{\odot})=151.74M_{\odot}$, with the restriction that $\frac{d}{dm}f_{\text{PPISN}}(m)=0$ here. The values for the coefficients are, then:
\begin{equation}
    \label{eqc3}
    \begin{pmatrix}a_0\\a_1\\a_2\\a_3\\a_4\end{pmatrix}=
    \begin{pmatrix}
        -3.38751\cdot10^2\\
        1.01736\cdot10^1\\
        -1.06666\cdot10^{-1}\\ 
        5.18494\cdot10^{-4}\\
        -9.74407\cdot10^{-7}
    \end{pmatrix}
    \text{ and }
    \begin{pmatrix}b_0\\b_1\\b_2\\b_3\\b_4\end{pmatrix}=
    \begin{pmatrix}
        -3.47418\cdot10^4\\
        8.53251\cdot10^2\\
        -7.84251\\
        3.20206\cdot10^{-2}\\
        -4.90175\cdot10^{-5}
    \end{pmatrix}
    \quad,
\end{equation}
this fit is valid for at least $100M_{\odot}\leq m\leq192M_{\odot}$. Now, in order to describe the failed SN population, apart from the BH island, we use the work of \citet{schneider}. Their results entail both a relation between ZAMS mass and CO core mass and a fit for the failed SN population, which depends on this relation. For the case A/B mass transfer, this comes down to:
\begin{equation}
    \label{eqc4}
    f_{\text{FSNe}}^{\text{AB}}(m)=15.832M_{\odot}\exp(0.058\left[\dfrac{3}{20}\dfrac{m}{M_{\odot}}+4\right])-14.784M_{\odot}\quad,
\end{equation}
where the part in square brackets is a linear fit to the CO core mass relation and the last term determines the upper border of the mass gap. Similarly, the failed SN function for case C mass transfer equals
\begin{equation}
    \label{eqc5}
    f_{\text{FSNe}}^{\text{C}}(m)=9.727M_{\odot}\exp(0.041\left[\dfrac{16}{31}\dfrac{m}{M_{\odot}}-\dfrac{128}{31}\right])+4.594M_{\odot}\quad.
\end{equation}
Eq. \ref{eqc4} and eq. \ref{eqc5} determine the upper border of the mass gap, at $22M_{\odot}$. Now, for the BH island we simply use a linear function which produces remnants ranging from $8M_{\odot}$ up to the lower border of the mass gap, at $14M_{\odot}$. These different functions are valid in ZAMS mass ranges which are similar to the values used by \citet{schneider} and \citet{ppisn}, although (small) deviations from these values should not influence the general shape of the remnant distribution. For case A/B mass transfer, this comes down to
\begin{equation}
    \label{eqc6}
    f_{\text{RN}}^{\text{AB}}(m)=\left\{\begin{matrix}
    0 &\text{for }m<30.5M_{\odot}\hfill\\
    1.2m-28.6M_{\odot} &\text{for }30.5M_{\odot}\leq m\leq35.5M_{\odot}\hfill\\
    0 &\text{for }35.5M_{\odot}<m<70M_{\odot}\hfill\\
    f_{\text{FSNe}}^{\text{AB}}(m)&\text{for }70M_{\odot}\leq m\leq100M_{\odot}\hfill\\
    f_{\text{PPISNe}}(m)&\text{for }100M_{\odot}<m\leq192M_{\odot}\hfill\\
    0 &\text{for }m>192M_{\odot}\hfill
    \end{matrix}\right.\quad,
\end{equation}
where we assume that not only PISNe but also regular successful SNe do not form remnants, since we are not interested in the distribution of neutron stars. The function for case C is similar to eq. \ref{eqc6}:
\begin{equation}
    \label{eqc7}
    f_{\text{RN}}^{\text{C}}(m)=\left\{\begin{matrix}   
    0 &\text{for }m<20.5M_{\odot}\hfill\\
    2.0m-33.0M_{\odot} &\text{for }20.5M_{\odot}\leq m\leq23.5M_{\odot}\hfill\\
    0 &\text{for }23.5M_{\odot}<m<35.5M_{\odot}\hfill\\
     f_{\text{FSNe}}^{\text{C}}(m)&\text{for }35.5M_{\odot}\leq m\leq58M_{\odot}\hfill\\
     f_{\text{PPISNe}}(m+42M_{\odot})&\text{for }58M_{\odot}<m\leq150M_{\odot}\hfill\\
     0 &\text{for }m>150M_{\odot}\hfill
    \end{matrix} \right.\quad,
\end{equation}
where the PPISN function is shifted to lower masses in order to connect with the failed SN function. Eq. \ref{eqc6} and eq. \ref{eqc7} give the curves which are shown in fig. \ref{fig2}. 

\section{Model uncertainties}
\label{app.D}
\begin{figure}[h]
    \resizebox{\hsize}{!}{\includegraphics{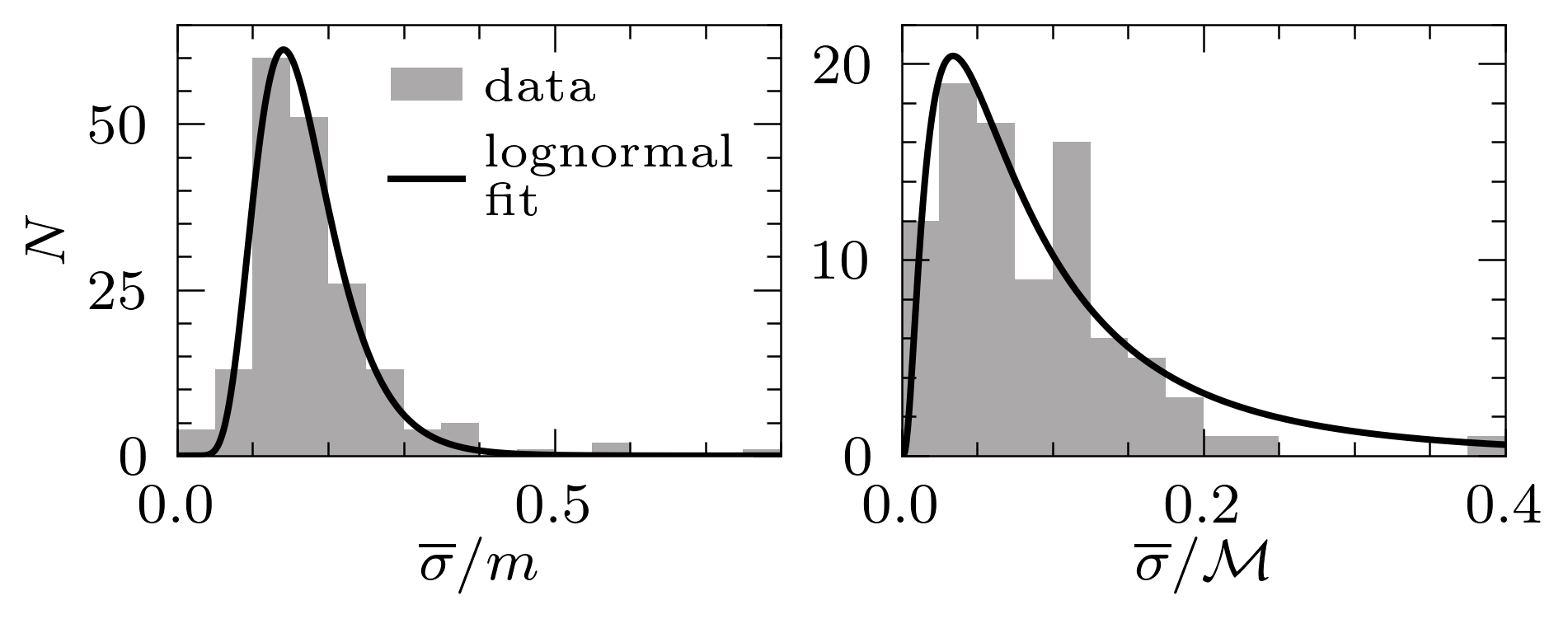}}
    \caption{The relative average standard deviations in the GW data \citep{LIGOopensource} for both individual masses (left) and chirp masses (right).}
    \label{figd1}
\end{figure}

\noindent In order to apply a Gaussian uncertainty to the results of our estimation in fig. \ref{fig4}, we make lognormal fits to the average relative standard deviations of the GW data \citep{LIGOopensource}. Fig. \ref{figd1} shows the data together with our fits. Applying this to our model, we pick a standard deviation randomly from the lognormal distributions and replace the result by a randomly chosen point from a Gaussian with the result as mean and the chosen standard deviation. However, the lognormal fits have a sharp enough peak that taking a single value for the relative standard deviations, e.g. at the maximum, would not influence the results significantly.
\section{Mass transfer efficiency and type}
\label{app.E}
\begin{figure}[h]
    \resizebox{\hsize}{!}{\includegraphics{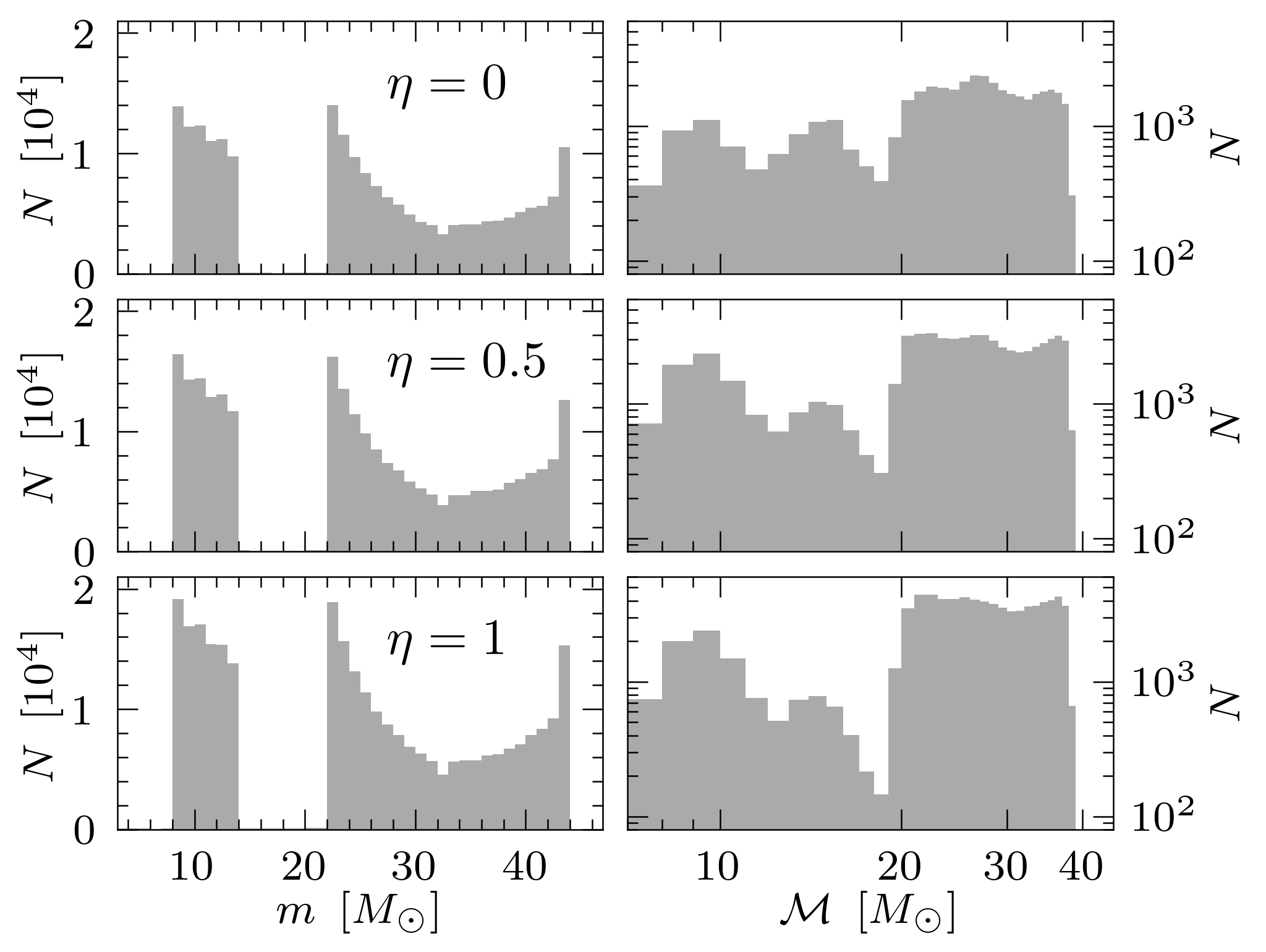}}
    \caption{Results of the estimation from fig. \ref{fig4}, for varying $\eta$ and $\zeta=0.5$.}
    \label{fige1}
\end{figure}

\begin{figure}[h]
    \resizebox{\hsize}{!}{\includegraphics{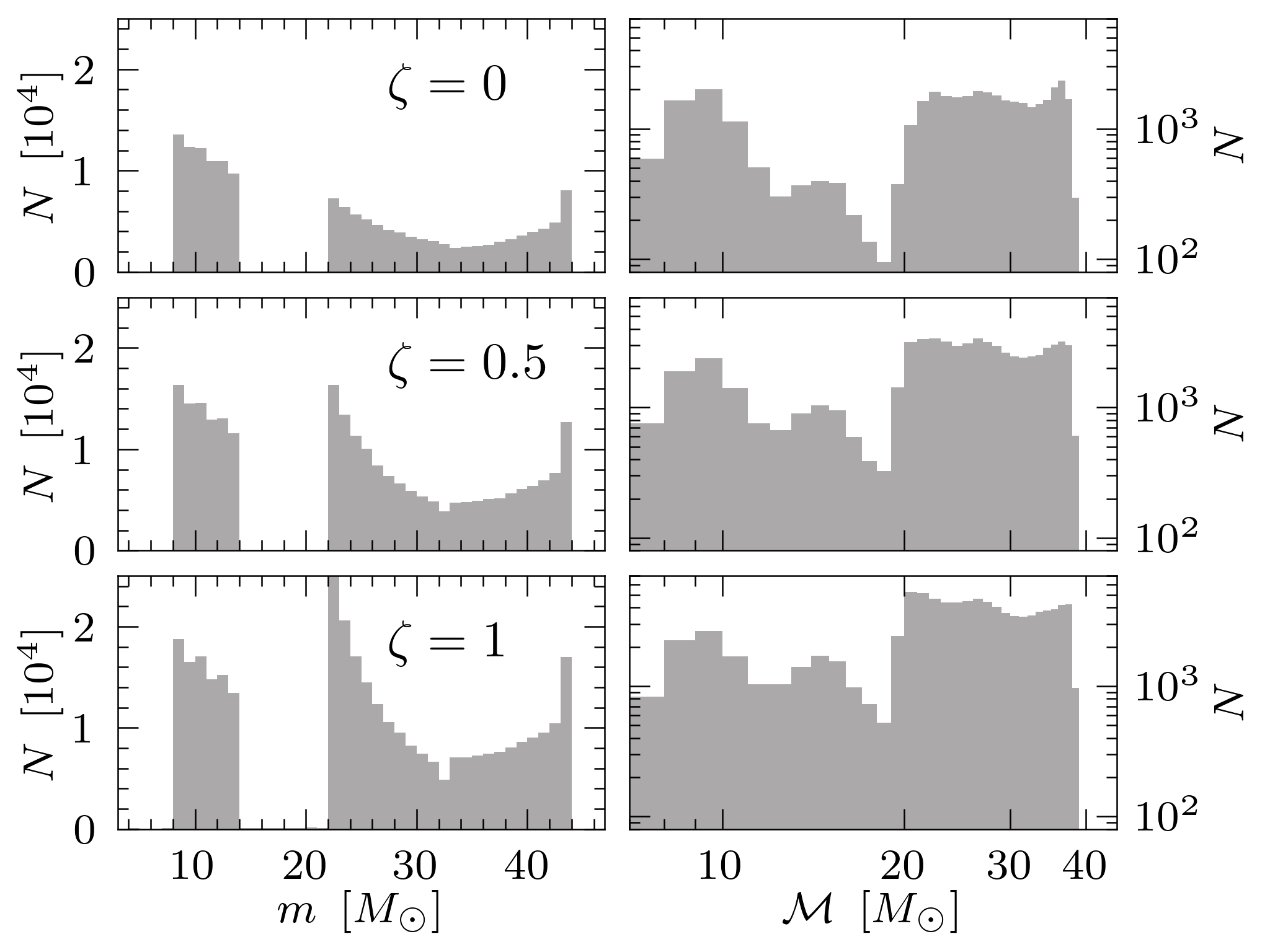}}
    \caption{Results of the estimation from fig. \ref{fig4}, for varying $\zeta$ and $\eta=0.5$.}
    \label{fige2}
\end{figure}

\noindent Fig. \ref{fige1} and fig. \ref{fige2} show the results of our estimation (fig. \ref{fig4}), but for varying $\eta$ and $\zeta$. The figures show that both parameters do not greatly influence the final mass distributions. The strongest influence of these factors concerns the peak at $m\approx22M_{\odot}$ and the gap at $\mathcal{M}\approx19M_{\odot}$, but these effects are not big enough to influence our conclusions.
\end{appendix}

\end{document}